\documentstyle[12pt]{article}

\begin{document}

\begin{center}
{\Large{\bf Numerical Modelling of Transport Processes in
Semiconductors} \\ [5mm]
E.P. Yukalova$^1$ and V.I. Yukalov$^2$} \\ [3mm]
{\it
$^1$Laboratory of Informational Technologies \\
Joint Institute for Nuclear Research, Dubna 141980, Russia \\ [2mm]

$^2$Bogolubov  Laboratory of Theoretical Physics \\
Joint Institute for Nuclear Research, Dubna 141980, Russia}

\end{center}

\vskip 2cm

\begin{abstract}

The peculiarities of electric current in semiconductors with nonuniform
distribution of charge carriers are studied. The semiclassical
drift-diffusion equations consisting of the continuity equations and the
Poisson equation are solved numerically using Rusanov finite-difference
scheme of third order. The different types of boundary conditions are
numerically investigated. It is shown that the stability of the Rusanov 
scheme for the problem considered is achieved with the Neumann type boundary
conditions. These conditions correspond to the absence of diffusion
through semiconductor surface. Special set of parameters is found under 
which a very interesting and unusual transient effect of negative current 
in nonuniform semiconductors appears. Different regimes of negative current 
are considered for realistic semiconductor materials.

\end{abstract}

\vskip 3cm


\vskip 1cm


\newpage

The study  of electric current in semiconductors is important for
describing and modelling semiconductor devices. The action of external
radiation can result in the formation of a nonuniform distribution of
charge carriers in a  semiconductor sample. The transport
processes in semiconductors with  essentially nonuniform distribution of
carriers can be quite specific. For instance, in a semiconductor sample
biased with an external constant voltage the electric current may turn
against the applied voltage [1,2]. Of course, it can happen only as a short
time fluctuation after which the current turns back becoming positive. So we
can talk about this only as about a transient effect. This unusual transient
effect of negative current is interesting by itself although there might
be some useful applications suggested. For example, it can be used for
measuring physical characteristics of semiconductor materials or for
constructing electronic controlling devices [3].

\vskip 2mm

Transport properties of semiconductors are usually described by the
semiclassical drift-diffusion equations [4]. In dimensionless units
we have the continuity equation
\begin{equation}
\label{1}
\frac{\partial\rho_i}{\partial t} + \mu_i\;
\frac{\partial}{\partial x}\; (\rho_i\; E) - D_i\;
\frac{\partial^2\rho_i}{\partial x^2} + \gamma_i\rho_i = 0 \; ,
\end{equation}
where $0\leq x\leq 1$, $t>0$, $i=1,2$, and the Poisson equation
\begin{equation}
\label{2}
\frac{\partial E}{\partial x} = 4\pi(\rho_1+\rho_2)\; .
\end{equation}
Here, $\rho_1=\rho_1(x,t)$ and $\rho_2=\rho_2(x,t)$ are charge densities;
$E=E(x,t)$, electric field. The functions $\rho_1,\; \rho_2,\; E$ are
unknown functions to be found from equations (1) and (2).

The considered semiconductor sample of length $L$ is characterized by the
following parameters: $\mu_i$, mobilities; $D_i$, diffusion coefficients;
$\gamma_i$, relaxation widths. Two types, ($i=1,2$), of charge carriers are
assumed: positive and negative, so that $\rho_1>0,\; \mu_1>0$, and
$\rho_2<0,\; \mu_2<0$. We consider the semiconductor sample of the
length $L=1$.

Initial distribution of charge carriers is modelled by the Gaussian form,
which e.g. corresponds to the distribution of charges in a semiconductor
irradiated by an ion beam. For $t=0$,
\begin{equation}
\label{3}
\rho_i(x,0) \equiv \frac{Q_i}{Z_i}\; \exp\left\{ -\;
\frac{(x-a_i)^2}{2b_i}\right\} \; , \qquad (0< a_i <1) \; ,
\end{equation}
where $Q_i=\int_0^1\rho_i(x,0)\; dx$ and $Z_i=\int_0^1\exp\left\{ -
\frac{(x-a_i)^2}{2b_i}\right\} dx$.
Note that the initial distribution for the electric field $E(x,0)$ is to be
found from equations (1) and (2), using (3) and given boundary conditions.

There are two possibilities of posing boundary conditions for functions
$\rho_i(x,t)$. The Dirichlet type,
$$
\rho_i(0,t) =\rho_i(1,t) = 0 \; , \qquad (i=1,2)\; ,
$$
which corresponds to the absence of the drift current $(j_i\equiv
\mu_i\rho_iE=0$) at the semiconductor surface, and the Neumann type,
$$
\left. \frac{\partial\rho_i}{\partial t} \right |_{x=0} =
\left. \frac{\partial\rho_i}{\partial t} \right |_{x=1} = 0 \; , \qquad
(i=1,2) \; ,
$$
which corresponds to the absence of diffusion through the semiconductor 
surface.

We considered both types of boundary conditions for $\rho_i$. It turned 
out that calculational procedure is less stable in the case of the Dirichlet 
boundary conditions because of the accumulation of charges at the 
semiconductor surfaces. As a result, discontinuities appear in the 
vicinity of the boundaries $x=0$ and $x=1$, which leads to numerical 
instability. In practice, it is also more difficult to arrange that 
the drift current through a semiconductor surface be absent. In what 
follows, we use the Neumann type of boundary conditions for the system 
(1) and (2).

The role of boundary conditions for the electric field $E(x,t)$ is played by
the voltage integral
$$
\int_0^1\; E(x,t)\; dx = 1 \; .
$$

Our aim is to study the peculiarity in the behaviour of the electric current 
through the semiconductor sample
\begin{equation}
\label{5}
J(t) = \int_0^1 j_{tot}(x,t)\; dx \; ,
\end{equation}
where
\begin{equation}
\label{6}
j_{tot}(x,t)=\left ( \mu_1 E - D_1\; \frac{\partial}{\partial x}\right )
\rho_1 +\left ( \mu_2 E - D_2\; \frac{\partial}{\partial x}\right ) \rho_2
+ \frac{1}{4\pi}\; \frac{\partial E}{\partial t} \; .
\end{equation}
We also are interested in the electric currents at the left ($x=0$) and
right ($x=1$) surfaces of the semiconductor, $J_0(t) \equiv j_{tot}(0,t)$ 
and $J_1(t)\equiv j_{tot}(1,t)$.
Let us note that $J(t)$ can be presented in the integral form
$$
J(t) = \int_0^1 \left [ \mu_1\rho_1(x,t) +\mu_2\rho_2(x,t)\right ] E(x,t)\;
dx +
$$
$$
+ D_1[\rho_1(0,t)-\rho_1(1,t)] + D_2[\rho_2(0,t)-\rho_2(1,t)] \; ,
$$
which is more convenient for numerical computation.

Numerical investigation showed that the known explicit finite difference
schemes of the second order are strongly unstable for the system (1) and (2).
Because of this, we resorted to the third-order accuracy finite-difference
scheme suggested by Rusanov [5] which proved to be very efficient for
discontinuous solutions. The calculational procedure for the Rusanov scheme
is constructed as follows. We consider the vector-function $u_m^{(n)}=\{
\rho_1(x_m,t_n),\rho_2(x_m,t_n),E_m(x_m,t_n)\}$, where $x_m=mh$ and
$t_n=n\tau$, $m=0,1,\ldots,M$; $n=0,1,\ldots$. The grid space step is $h=1/M$.
The time step $\tau$ is chosen so that the calculational procedure 
be stable. The parameter $\sigma$ in the Rusanov scheme, responsible for
the stability of calculations, is $\sigma=0.45$ for $\tau/h=0.125$. We will
not present the computational scheme explicitly since it is quite
cumbersome. Let us better dwell upon the approximation of boundary conditions
for $\rho_i(x,t)$ ($i=1,2$) and for $E(x,t)$. Numerical modelling
showed that the Neumann type of boundary conditions for the charge carriers
are to be approximated with the third order accuracy for the time layers
$t=t_n=n\tau$ and auxiliary time layers $t=t_n^{(2)}=t_n+2\tau/3$ at the
integer boundary nodes:
$$
\rho_i(x_{M},t) = \frac{1}{3}\; \left [ 4\rho_i(x_{M-1},t)-\rho_i(x_{M-2},t)
\right ]\; , \qquad \rho_i(x_0,t) = \frac{1}{3}\; \left [ 
4\rho_i(x_1,t) -\rho_i(x_2,t)\right ] \; ,
$$
and with the second order accuracy for the half-integer boundary nodes
at the auxiliary time layer $t=t_n^{(1)}=n\tau+\tau/3$:
$$
\rho_i(x_{M+1/2},t) = \rho_i(x_{M-1/2},t) \; , \qquad
\rho_i(x_{-1/2},t) = \rho_i(x_{1/2},t) \; ,
$$
where $x_{m+1/2}=mh+1/2$, $m=0,1,\ldots,M$, $n=1,2,\ldots$.

To approximate the boundary conditions for the electric field $E(x,t)$,
given in the integral form, it was necessary to rewrite it in the form
convenient for calculations:
$$
E(x,t) = 1 +4\pi\left [ Q(x,t) -\int_0^1\; Q(x,t)\; dx \right ] \; ,
\qquad Q(x,t) = \int_0^x [\rho_1(y,t)+\rho_2(y,t)]\; dy \; .
$$
Then $E(0,t)$ and $E(1,t)$ are given by the above formulas for $x=0$ and
$x=1$. At each time layer $t_n=n\tau$, we checked whether the equality 
$\int_0^1 E(x,t)dx=1$ holds true.

We have accomplished numerical calculations for different sets of parameters
$a_i$, $b_i$, $\mu_i$, $D_i$, $\gamma_i$, and $Q_i$. It turned out that
for some values of parameters there exists an unusual transient effect when
the electric current turns against the applied voltage. This effect of 
{\it negative current} occurs on the whole manifold of parameters. Here we 
illustrate it for the case of a unipolar semiconductor, when $Q_1\equiv Q>0$
and $Q_2=0$. Other parameters are set as $\mu_1=1,\; \mu_2=0,\; D_i\ll 1,\;
\gamma_i\ll 1$, and the notation $a_1\equiv a$, $b_1\equiv b$ is used.

Fig. 1 shows the total electric current (4) through the semiconductor sample, 
as a function of time, for the injected electric charge $Q=0.5$, the
initial Gaussian distribution width $b=0.1$, and for different locations of 
the peak of this charge distribution, $a$. As is seen, the negative electric 
current happens at the initial stage of the process for $a<0.5$, such as
$a=0.1$ and $a=0.25$, but does not exist for $a\geq 0.5$.

Fig. 2 presents the total electric current versus time for $Q=3$ and $b=0.5$, 
with varying locations $a$. Again, the negative current appears for $a<0.5$
but does not occur for $a>0.5$. More accurately, the conditions on $a$ and 
$Q$, sufficient for the occurrence of the negative current, are
$$
a< \frac{1}{2}\; - \; \frac{1}{4\pi Q} \; , \qquad
Q>\frac{1}{2\pi} \; .
$$

The existence of the transient effect of negative electric current is the
most interesting result of the numerical modelling we have accomplished for
semiconductors.

\newpage

\begin{center}

{\large{\bf Figure captions}}

\end{center}

{\bf Fig. 1}. Total electric current as a function of time for $Q=0.5$,
$b=0.1$ and different locations for the peak of the initial charge-carriers
distribution: $a=0.1$ (1 curve, solid line); $a=0.25$ (2 curve, long-dashed
line); $a=0.5$ (3 curve, short-dashed line); and $a=0.75$ (4 curve, dotted
line).

\vskip 5mm
{\bf Fig. 2}. Electric current through the semiconductor sample versus 
time for $Q=3$, $b=0.5$, and varying initial charge locations $a$, which take 
the same sequence of values as in Fig. 1.

\end{document}